\documentclass{article}
\usepackage[margin=1in]{geometry}

\usepackage{graphicx}
\usepackage{amsmath}
\usepackage[lofdepth,lotdepth]{subfig}
\usepackage{enumitem}
\setlist[itemize]{leftmargin=*}

\usepackage{eso-pic}

\title{Towards Malicious address identification in Bitcoin}

\author{
        Deepesh Chaudhari, Rachit Agarwal,
        Sandeep Kumar Shukla \\
        CSE, IIT Kanpur, India\\
        Email: \{deepesh,rachitag,sandeeps\}@iitk.ac.in
}
\date{}

\providecommand{\keywords}[1]
{
  \small	
  \textbf{\textit{Keywords---}} #1
}

\begin{document}

\AddToShipoutPictureBG*{%
  \AtPageUpperLeft{%
    \setlength\unitlength{1in}%
    \hspace*{\dimexpr0.5\paperwidth\relax}
    \makebox(0,-0.75)[c]{\Large Author's copy}%
}}

\maketitle

\begin{abstract} 
The temporal aspect of blockchain transactions enables us to study the address's behavior and detect if it is involved in any illicit activity. However, due to the concept of change addresses (used to thwart replay attacks), temporal aspects are not directly applicable in the Bitcoin blockchain. Several pre-processing steps should be performed before such temporal aspects are utilized. We are motivated to study the Bitcoin transaction network and use the temporal features such as burst, attractiveness, and inter-event time along with several graph-based properties such as the degree of node and clustering coefficient to validate the applicability of already existing approaches known for other cryptocurrency blockchains on the Bitcoin blockchain. We generate the temporal and non-temporal feature set and train the Machine Learning (ML) algorithm over different temporal granularities to validate the state-of-the-art methods. We study the behavior of the addresses over different time granularities of the dataset. We identify that after applying change-address clustering, in Bitcoin, existing temporal features can be extracted and ML approaches can be applied.
A comparative analysis of results show that the behavior of addresses in Ethereum and Bitcoin is similar with respect to in-degree, out-degree and inter-event time. Further, we identify 3 suspects that showed malicious behavior across different temporal granularities. These suspects are not marked as malicious in Bitcoin.

\end{abstract}

\vspace{-0.3cm}
\section{Introduction}

Blockchain technology was first introduced with the introduction of Bitcoin by Satoshi Nakamoto in 2008~\cite{Satoshi2008} to increase the trust and transparency of users by shifting the controlling power from central authority to a decentralized network. This also allowed the users to transact anonymously and without any fear of being traced. In the Bitcoin blockchain, users transact using cryptocurrency called Bitcoin (BTC). Due to these features, cyber-criminals quickly adopted Bitcoin to perform malicious activities such as \textit{ransomware attacks}, \textit{money laundering}, \textit{gambling}, \textit{ponzi schemes}, and \textit{phishing}~\cite{darknet2017}.  With increased adoption of blockchain by different industries, such activities have grown to cause more financial losses. In~\cite{ChainAnalysis2021}, the authors identify that in the year 2020, illicit entities made $\approx$\$10 Billion worth of transactions.

With more and more security threats appearing with the use of blockchains, we ask \textit{whether malicious user's addresses can be detected in a blockchain using approaches such as machine learning (ML)}? Many state-of-the-art algorithms aim to detect malicious addresses in Bitcoin, Ethereum, and other permissionless blockchains. These algorithms use either \textit{(a)} graph-based techniques \cite{zola2019ieee,shikar2020} or \textit{(b)} ML algorithms~\cite{pham2017anomaly,monamo2016,Chen2018WWW,Bartoletti2018}. To perform ML-based detection, it is important to understand how malicious activities are different from benign activities. Temporal aspects inherent in the users's behavior eases such understanding. Nonetheless, such temporal aspects are rarely used by the state-of-the-art approaches and no feasibility study is performed to identify whether the features apply to the blockchain or not. For example, in the Bitcoin blockchain, to receive the remaining balance after a transaction, a new address of the address is generated (also called as change address). Such aspects allow a user to hold multiple addresses. In such a scenario, the application of temporal features such as in-degree would be instantaneous. 

Hence, the research questions (RQ) that we ask are: \textbf{\textit{(RQ1)}} Can we detect malicious addresses in Bitcoin using ML techniques and consider the temporal behavior? \textbf{\textit{(RQ2)}} Are the features identified in state-of-the-art approaches targeting blockchains such as Ethereum applicable in Bitcoin? As there are change addresses in Bitcoin, \textbf{\textit{(RQ3)}} What changes occur if we cluster the addresses? Further, once addresses are clustered \textbf{\textit{(RQ4)}} Does behavioral changes exist in Bitcoin? 
To answer these research questions, we create two graphs (\textbf{user/address} and \textbf{transaction} graph) from the Bitcoin transaction data, extract required features from these two graphs, and apply ML techniques to detect addresses we suspect to be malicious. We find that features such as \textit{degree burst} (an abrupt rise in the degree of an address), \textit{balance burst} (an abrupt increase in the balance of an address), and \textit{attractiveness} (probability of interaction with the same address that the address interacted with before~\cite{shikar2020}) are not directly applicable in Bitcoin because of the concept of change address. Note that for this work we focus on those malicious entities for which we have ground truth data. Our approach is agnostic to the type of malicious activity.

Change addresses are used to protect the privacy and increase the user's anonymity. Thus, features like degree burst, balance burst, and attractiveness that use temporal aspects of transactions are spread over different addresses. Calculating these features directly from the Bitcoin transaction data does not provide any helpful information about the address's behavior. To identify the addresses potentially held by the same user, in~\cite{address_clustering_heuristics}, the authors develop two heuristic approaches \textit{(a)} \textbf{multi-input} heuristics (based on the assumption that if a transaction uses more than one address as an input, then they must be created and controlled by the same user) and \textit{(b)} \textbf{change address} heuristic (addresses that are created by the same user to keep the change amount (remaining BTCs)). Note that there are other approaches (described in Section~\ref{sec:Relatedwork}) that cluster the addresses, but we apply the multi-input and change address heuristics to cluster the Bitcoin addresses in this work as they provide the best results~\cite{address_clustering_heuristics}. After using the heuristics, we find that the total number of users that transacted during the considered period reduces considerably.

Next, we segment the data using different temporal granularities ($T_G$) to understand the changes in the behavior. These temporal granularities include 15Days and 1Month granularities, i.e., $T_G$=\{15Days, 1Month\}. Each data segment (SD) is mutually exclusive in these granularities and comprises of transactions in a given particular period defined by that segment. On these SDs, we first apply unsupervised ML (K-Means) to identify malicious addresses. For comparison, we then apply heuristics-based change address clustering on each of the SDs and apply unsupervised ML. 

In summary, our main contributions are:
\begin{itemize}
    \item \textbf{\textit{Validation of the state-of-the-art methods}}: We apply the state-of-the-art ML approaches to detect malicious addresses in the Bitcoin. We find that the features present in other works are applicable to Bitcoin, only after applying change-address heuristics. Using the Bitcoin transaction data, we verify that temporal features such as in-degree, out-degree and inter-event time follow either \textit{power-law} distribution or its variations. Thus, proving existence of burst. We also find that using temporal features more suspects are detected.
    
    \item \textbf{\textit{Comparative analysis}}:  Using KL divergence, we find that in Bitcoin, behavior of addresses in similar to those in Ethereum with respect to features such as in-degree, out-degree and inter-event time. 
    
    \item \textbf{\textit{Behavior analysis}}: We perform the behavior analysis on Bitcoin addresses using the temporal and non-temporal features extracted from their transactions.
    Using temporal features we find a total of \textbf{\textit{3,273}} and \textbf{\textit{19,712}} addresses that change their behavior in the considered two different time granularities. Whereas, using non-temporal features we find a total of \textbf{\textit{868}} and \textbf{\textit{159}} addresses that change their behavior in the two different time granularities.
\end{itemize}

This paper is organized as follows. In Section~\ref{sec:Relatedwork}, we present the state-of-the-art techniques for identifying malicious addresses and compare them. In Section~\ref{sec:methodology}, we give a detailed description of our methodology. This section is followed by an in-depth evaluation of the results in Section~\ref{sec:Evaluation} that includes a description of the data. We finally conclude in Section \ref{sec:Conclusion} and also provide prospective future directions.

\vspace{-0.3cm}
\section{Related work}\label{sec:Relatedwork}

In this section, we first present different state-of-the-art approaches towards de-anonymizing the Bitcoin address and then present those approaches that tackle the identification of malicious addresses in the Bitcoin blockchain. Further, we also survey the state-of-the-art approaches from a more general perspective, especially those that deal with permissionless blockchains and claim that their approach applies to any permissionless blockchains. Most of the approaches towards identifying malicious addresses, either in Bitcoin or in permissionless blockchains, are ML-oriented, where the feature vector is based on graph and transaction analysis. Note that although there are lot of state-of-the-art approaches that use temporal aspects, we do not focus on them as they predict price of Bitcoin and not illicit entities.

\vspace{-0.3cm}
\subsection{De-anonymizing the Bitcoin addresses}\label{sec:address_clustering}

Assume a transaction $T_x$ is performed between input addresses (senders represented by the set $inputs(T_x)$) and output addresses (receivers represented by the set $outputs(T_x)$).

In~\cite{Heuristic_paper_bitcoin, Multi-input_Reid2013}, the authors introduced the multi-input heuristics which assumes that a single user controls all the $input(T_x)$ of the $T_x$. A sender needs to sign all the inputs using the private keys to prove that the user owns the BTCs present in those addresses to make a valid transaction. Hence, the user that created the transaction must control all those private keys associated to $input(T_x)$. For a transaction $T_x$, consider the $C_{T_x}$ to be the cluster of addresses controlled by one user. Then, using this method, $C_{T_x}$=$inputs(T_x)$.

In Bitcoin, when a user makes a transaction, he needs to transfer all the BTCs he has to receiver's address. If the user wants to transfer only some BTCs he has to another user's address, then he must create a new address and send the remaining BTCs to that address. Since, the user controls the input addresses and this change address, these addresses should be clustered together. In~\cite{Heuristic_paper_bitcoin}, the authors proposed a heuristics-based method to identify such addresses. An address ${o_j}(T_x) \in outputs(T_x)$ is said to be a ``change'' address if:

\begin{enumerate}
    \item The address ${o_j}(T_x)$ appears first time.
    \item Transaction $T_x$ is not a ``coin-generate'' transaction.
    \item There must be no common address between output addresses and input addresses (No \textbf{self-change address}).
    \item The first condition is only satisfied for ${o_j}(T_x)$ and not for other addresses in $outputs(T_x)$.
\end{enumerate}    

Thus, combining the two methods, for a transaction $T_x$, ${C_{T_x}}$=$inputs(T_x)\cup \{o_j(T_x)\}$.

In~\cite{Multi-input_Reid2013}, the authors proposed two more conditions to capture the ``\textit{no}'' change address scenario. A $T_x$ contains no change address if there is an address in the output such that:

\begin{itemize}
    \item [5)] It has already received only one input.
    \item [6)] It has appeared before in a self-change transaction.
\end{itemize}

In~\cite{address_clustering_heuristics}, the authors introduced a  \textbf{\textit{value-based heuristics}} method as an extension to the above conditions. They assumed that users try to minimize the size of the transaction by reducing the number of inputs and outputs. An $o_j(T_x)$ is expected to be a change address if the $T_x$ contains only one output with a value less than its inputs. If the change amount was more than any input, then we could ignore the input. Further, in~\cite{address_clustering_heuristics}, the authors proposed a \textbf{\textit{growth-based heuristics}} method and showed that clusters usually increase steadily. It's rare to see the amalgamation of two big clusters by a new transaction and might indicate a false positive from one of the heuristics. 

In~\cite{Zhang2020Heuristic}, the authors, besides using multi-input and change-address  heuristics methods, use an address reuse-based heuristics method. Here, instead of previously defined condition `4', the authors proposed:

\begin{itemize}
     \item[4')] Output addresses except for $o_{j}(T_x)$ are reused as an output address in some later transactions.
\end{itemize}

In~\cite{address_clustering_heuristics}, the authors also  compared the heuristics methods to know which method provides better results. They found that multi-input together with change address heuristics gives a more reliable cluster. 
\vspace{-0.2cm}
\subsection{ML Based detection approaches}

In~\cite{pham2017anomaly}, the authors used unsupervised learning to detect malicious addresses in Bitcoin. They used K-Means as a baseline method and calculate the \textit{local outlier factor (LOF)}. For the feature extraction, they analyzed the Bitcoin transaction data and created two graphs, \textbf{user} and \textbf{transaction} graphs. They extracted features corresponding to graph properties such as inter-event time, balance, and activity period. Using transactions of 6.3 Million addresses, they found that for $K$=8, K-Means provided optimal results for both the user and the transaction graph. However, they did not mention how they create the user graph. 

In a similar work, in~\cite{monamo2016}, the authors compared the results of K-Means and trimmed K-Means on the Bitcoin transaction data of 1 Million addresses. However, they limited their study to graph-based and balance-based features. They found that both trimmed K-Means and K-Means provided optimal results when $K$=8 and thus also validated~\cite{pham2017anomaly}. Nonetheless, they found that trimmed K-Means provided better results with an improved detection rate.

In~\cite{Bartoletti2018}, the authors used supervised ML algorithms to detect addresses involved in \textit{Ponzi schemes} in Bitcoin. A Ponzi scheme is an investment fraud that generates returns to existing investors with funds raised from new investors. Here, the authors de-anonymized the Bitcoin addresses using the above mentioned heuristics methods. They, used a rule-based learner called RIPPER, Bayes Network, and Random-Forest classifiers on the features (such as graph properties, inter-event time, balance, and activity period) extracted from 6432 Ponzi scheme related addresses. They found that both RIPPER and Random-Forest classifier achieved high accuracy. However, the method only focused on Ponzi schemes and did not detect addresses involved in any other malicious activity.   

In~\cite{zola2019ieee}, the authors studied transactions between entities (organizations or people with multiple addresses) to attack Bitcoin address anonymity. They present a novel \textit{cascading machine learning} approach that requires only a few features (graph-based, balanced-based, and transaction fee-based) extracted from transactions of 1000 Million addresses present in Bitcoin. Cascading, used to enrich an entity's information with data from previous classifications, considerably improves the multi-class classification performance for classes such as benign and malicious. As an initial step, the authors performed the classification of addresses into entities. From the data associated with these entities, they extract features and finally apply ML (especially Adaboost, Random-Forest, and Gradient Boosting models) to detect malicious addresses. For classification of addresses into entities, the authors used \textit{C\_address}, \textit{C\_motif1}, and \textit{C\_motif2}\footnote{\textit{N-motifs} are frequently occurring sub-graphs of a network. In~\cite{zola2019ieee}, the authors define \textit{N-motif} as a sub-graph of maximum \textit{2*N} path length between two entities, where transactions are also considered as vertices.}.
Here, C\_address contains features such as the number of transactions related to a Bitcoin address. C\_motif1 contains the information such as the number of transactions extracted from the \textit{1\_motif} graph  while C\_motif2 contains information such as the number of transactions gathered from the \textit{2\_motif} graph. Overall, the Gradient Boosting model provided the best classification scores when C\_address, C\_motif1, and C\_motif2 are used together.

In~\cite{regulating_cryptocurrencies_supervised22019}, the authors used supervised learning to classify 957 Bitcoin addresses that performed 385 Million transactions. They used heuristics-based algorithm first to cluster the Bitcoin addresses. Then, they used the clustered data to train the ML model and classify the addresses into 12 categories (exchange,  merchant-services, mixing, gambling, personal-wallet, mining-pool, ransomware, hosted-wallet, scam, darknet-market, stolen-Bitcoins, other). They used various supervised learning algorithms and found that Gradient Boosting gave the best result with a 79.64\% F1-Score. However, the authors do not comment on the bias induced by the address clustering heuristics while identifying malicious addresses and perform oversampling using SMOTE. On the contrary our work does not oversample to achieve better results.

In~\cite{BitcoinHeist2020}, the authors introduced \textit{Topological Data Analysis} (TDA) to train the ML model to detect ransomware-associated addresses in Bitcoin. The concept of TDA is to extract pattern hidden inside the data. The authors found that TDA performs best amongst different ML algorithms. They trained the algorithm using 6 features (total incoming balance, address from which the address has received BTC, sum of fraction of coins that reach the address, longest path to that address, number of addresses that have a path to the address, and number of address that can reach the address via multiple paths).

Given such state-of-the-art techniques to identify malicious addresses in Bitcoin, in~\cite{shikar2020}, the authors identified that these above-mentioned state-of-the-art approaches do not use temporal aspects to identify malicious addresses. Thus, they analyzed the temporal behavior of transactions present in the blockchain using temporal graphs and introduce new temporal features (such as burst and attractiveness) to detect malicious addresses. They studied the behavior of 700,000 addresses and applied supervised and unsupervised learning approaches. They found that ExtraTreesClassifier performed the best with balanced accuracy $>$87\%. While for the unsupervised learning case, they found that K-Means with $K$=10 performed the best. Although the authors validated their work on Ethereum transaction data, they claimed that their features are applicable to all the permissionless blockchains. For such a generic claim, where different permissionless blockchains have different architecture and implementations, such aspects need to be validated on a wider set of permissionless blockchains. 

In summary, all the state-of-the-art approaches have bias concerning the available ground truth. Except for~\cite{zola2019ieee,Bartoletti2018}, all the approaches consider all the addresses associated with malicious activities under one class. As malicious activity-specific data is hard to find for Bitcoin, we also focus on identifying malicious addresses without associating them with any malicious activity. However, in Ethereum, such information is available, which authors in~\cite{tanmayadversarial2021} leveraged to study how the ML approaches perform with respect to different activities. However, as we do not focus on Ethereum, we refrain from explaining them in this work and only focus on Bitcoin.
\vspace{-0.2cm}
\section{Methodology}\label{sec:methodology}

In this section, we discuss an outline of our methodology and the process pipeline. Here, we first describe how we pre-process transaction data to convert it. Then, we discuss the feature extraction, data configurations, and the ML techniques we use in our work.
\vspace{-0.2cm}
\subsection{Pre-processing and Graph generation}\label{methodology:pre-process}

A collection of Bitcoin transactions represents a temporal-directed social interaction graph. From this, we extract \textit{\textbf{Aggregated graph/Transaction graph}} (${AG^t(V^t,E^t)}$) and we convert it into \textit{\textbf{User graph}} (${UG^t(V^t,E^t)}$) to extract true nature of features such as indegree and outdegree. Here $t$ is defined as block number in our case. Note that these graphs are temporal as edges or interactions are intermittent causing graph properties to change over time.  In ${AG^t(V^t,E^t)}$, $V^t$ is a set of two types of addresses, one that belong to users and another to transactions at a time $t$. Every transaction node input and output is to addresses that belong to users. On the other hand, in ${UG^t(V^t, E^t)}$, $V^t$ is a set of addresses that belong to users at a time $t$, and $E^t$ is the set of edges or transactions between these addresses at a time $t$. To create these graphs, we neglect those transactions in which users transact using \textit{op\_return} or transact with \textit{script hash} because output addresses are not available in such cases.

In Bitcoin, due to the concept of change address, each user has multiple addresses. Such aspects make it hard to extract the temporal behavior of the user. To know what addresses belong to which users, many studies use heuristics-based approaches to cluster addresses.  We also apply the heuristics approaches mentioned in Section~\ref{sec:address_clustering} to identify addresses held by a user to both the user and aggregated graph.

\begin{figure}
    \vspace{-0.4cm}
    \centering
    \subfloat[15Days.][15Days.]{
        \includegraphics[width=0.7\columnwidth]{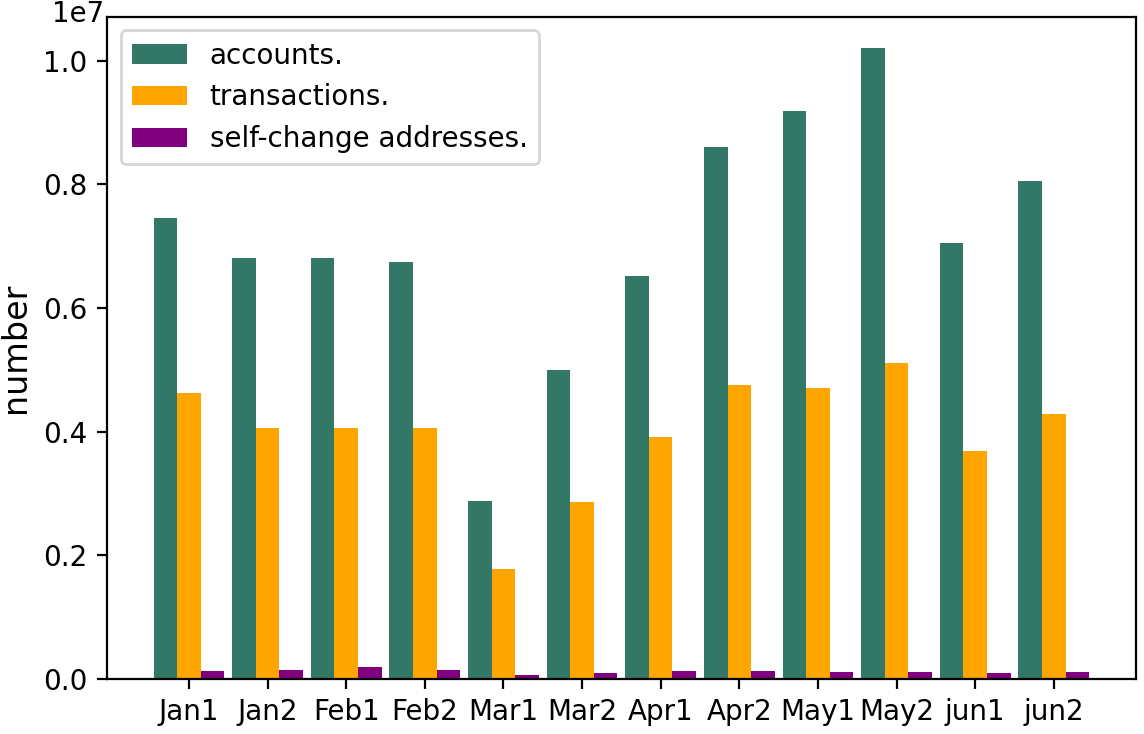}
        \label{fig:number_txns_add_15}
    }\\
    \subfloat[1Month.][1Month.]{
        \includegraphics[width=0.7\columnwidth]{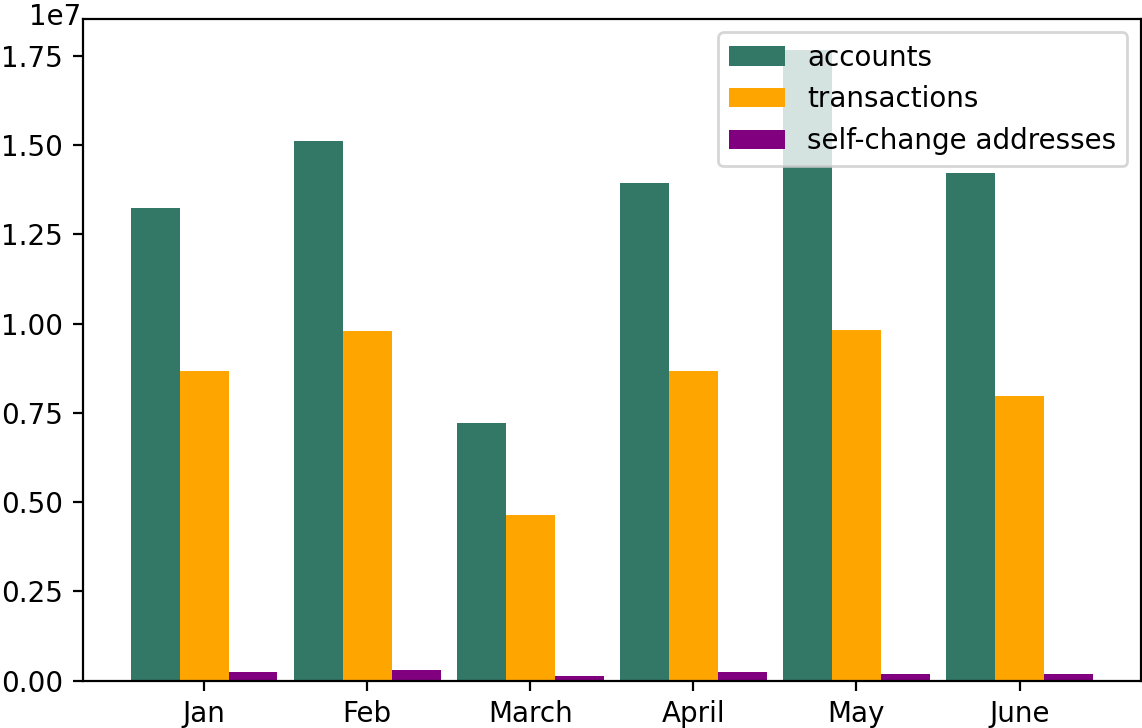}
        \label{fig:number_txns_add_30}
    }
    \caption{Distribution of transactions, addresses, and self-change addresses when considering different temporal granularities. Here, `Month*' represent whether it is first 15Days or second 15Days of the month.} 
    \label{fig:tempGranularityDistib} 
    \vspace{-0.3cm}
\end{figure}

\vspace{-0.2cm}
\subsection{Features Extraction}\label{methodology:dataframeandfeatures}

We generate our training and test dataset from ${UG^t(V^t, E^t)}$ and ${AG^t(V^t, E^t)}$ and extract two types of features: non-temporal features and temporal features as introduced in~\cite{shikar2020}. We discuss the non-temporal and temporal features next.

\subsubsection{Non-Temporal features}\label{sub:standard_dataset}
are those that are independent of time. We classify non-temporal features into those that are based on graph properties and balance. 

\begin{itemize}
    \item \textbf{Graph properties based features}:
     The graph properties are one of the best ways to characterize addresses. For example, \textit{in-degree} of an address shows the number of transactions the address has performed as a receiver, and \textit{unique in-degree} indicates the number of transactions that address performed as a receiver with unique addresses. Most malicious addresses receive BTCs from different addresses and have a higher number of transactions than benign addresses~\cite{shikar2020}. This makes in-degree and unique in-degree an important feature to differentiate between malicious and benign addresses. Similarly, when an address acts as a sender, \textit{out-degree} and \textit{unique out-degree} features are important. Most benign addresses prefer to perform transactions with already known addresses and pose a relatively less unique out-degree than malicious addresses for whom the neighborhood is rather less stable. This also impacts the \textit{Clustering-Coefficient} feature (which defines how tightly addresses are coupled with the neighbor addresses~\cite{Watts1998}). Nonetheless, we also include the total number of transactions in which an address is involved as \textit{total degree} and the total number of transactions performed with unique addresses as the \textit{unique total degree}.

    \item \textbf{Balance-based feature}: The amount of funds transferred/associated with a transaction plays an important role in determining whether an address is malicious or not. Malicious addresses at times transact large amounts as compared to benign addresses. Thus, features such as maximum payment received (\textit{max\_in\_payment}), maximum payment sent (\textit{max\_out\_payment}), the total amount received (\textit{in\_balance}), total amount sent (\textit{out\_balance}), and the balance of the address (\textit{txn\_sum}) are essential features. Nonetheless, most malicious addresses are less active as compared to benign addresses. Thus, features such as when the address first transacted (\textit{first\_txn\_block}), last transacted (\textit{last\_txn\_block}), and how long an address was active are essential features towards classifying an address as malicious.
\end{itemize}

\subsubsection{Temporal Features}\label{sub:temporal_dataset}

Due to the temporal nature, graph properties evolve and are not static. It is essential that the properties are not studied as static but studied as dynamic properties. One should extract features considering time to represent them better. In~\cite{shikar2020}, the authors identified features such as \textit{inter-event time} (time interval between two transactions of an address), \textit{attractiveness} (the probability of an address interacting with previously unknown address), and \textit{bursty} (abnormal rise and fall in the value of the property) behavior of \textit{in-degree}, \textit{out-degree}, and \textit{balance} as important features that distinguish malicious and benign addresses. As these features are temporal, we use time-series analysis to extract top 3 representative features of the above mentioned features using \textit{tsfresh}~\cite{christ2018}. Using such features, we also validate their applicability in Bitcoin to identify malicious addresses.

\vspace{-0.3cm}
\subsection{Data Configuration}\label{methodology:config}

We segment the data into several sub-datasets ($SD$) of different time granularities ($T_g$) to study the behavior of an address. Due to the limited availability of computational resources, we currently validate our approach on only 2 temporal granularities: 15Days and 1Month. Here, each granularity has $n$ different $SD$s that are mutually exclusive. For each $SD$, in each temporal granularity, we further generate 3 different datasets ($SD^{i,j}(T_g)$ where $i$ is the $i^{th}$ SD in a $T_g$ and $j\in \{1,2,3\}$). Here, $SD^{i,1}(T_g)$ is the dataset with only non-temporal features, $SD^{i,2}(T_g)$ is the dataset generated after clustering addresses and contains only non-temporal features, and $SD^{i,3}(T_g)$ is the dataset generated after clustering addresses and contains both temporal and non-temporal features. Note that $SD^{i,j}(T_g)\in SD^j(T_g)$.
\vspace{-0.2cm}
\subsection{Machine Learning}\label{methodology:machinelearning}

After generating different SDs mentioned in Section~\ref{methodology:config}, we apply results (K-Means with $K$=10) of~\cite{shikar2020} for unsupervised learning and validate it. Other supervised approaches like DNN or ensamble based methods although can provide better results but are out of the scope of this work. After clustering, we take the cluster with the most number of malicious addresses and compute their cosine similarity (CS) with other benign addresses in the corresponding cluster. CS is one of the most widely used measure that was also used in~\cite{shikar2020}. We thus use the same method as we want to validate the approach. Such approach allows us to be within our computational constraints and label only a few addresses as potential suspects. To find benign addresses with high cosine similarity with malicious addresses, we study the variation of the number of benign addresses tagged as malicious while changing the $\epsilon$ parameter. Here, $\epsilon$ is the threshold that identifies which benign addresses should be marked as suspected to be malicious. Formally, we mark benign addresses if equation~\ref{eq:similarity} is satisfied. We evaluate for $\epsilon\in$[0, 20].  
\begin{equation}\label{eq:similarity}
    k^{label} = 
    \begin{cases}
        \textit{malicious} ,& \text{for any } l\in X: \text{ 1}-\textit{CS}(k,l) \leq 10^{-\epsilon} \\
        
        \textit{benign},         & \text{otherwise}
    \end{cases}
\end{equation}

Here, $k$ is the benign address and $X$ is the set of malicious addresses. After identifying suspects in all the $SD$s, we calculate the probability $p^k$ of a particular address $k$ to be malicious in a given $T_g$ as stated in~\cite{shikar2020}. Here, $p^k$=$\frac{\sum_{i\in SD^{j}(T_g)}{M^k_i}} {n^k}$ where, $M^k_i$ depicts whether the address $k$ was identified as suspect in the $i^{th}$ sub-dataset ($SD^{i,j}(T_g)$) using $j$ features at temporal granularity $T_g$, and ${n^k}$ is the total number of SDs in which address $k$ transacted. 
\vspace{-0.2cm}
\section{Evaluation: Data and Results Analysis}\label{sec:Evaluation}

We evaluate the effectiveness of the state-of-the-art method using Bitcoin's transaction data. We set up a Bitcoin node\footnote{https://Bitcoin.org/en/full-node} and fetch transactions using Bitcoin RPC API\footnote{https://developer.Bitcoin.org/reference/rpc/}. Note that the APIs do not provide sensitive information about the users (such as the name, date of birth, and account type). Although as the address's hash is publicly available, one can look up the associated information using any Blockchain explorer such as Blockchain.com\footnote{https://www.blockchain.com/}. We perform all our evaluations using Python3 and scikit-learn.

\vspace{-0.3cm}
\subsection{Data}

In Bitcoin, more than 650 Million transactions are performed till June 2021~\cite{txn_in_bitcoin}, with the highest being 400,000 transactions per day in January 2021~\cite{highest_txn_in_bitcoin}. On average, in Bitcoin, a new block is created every 10 minutes. For our experiments, we choose Bitcoin transaction data from $1^{th}$ January 2020 till $30^{th}$ June 2020 due to computational constraints. During this period, 60,963,231 unique addresses performed 49,244,236 transactions. Figure~\ref{fig:tempGranularityDistib} show the number of transactions performed and the number of addresses that transacted at different periods that we consider. We see a dip in the number of transactions and the number of addresses that performed the transactions in March. This could be due to the rise of the COVID-19 pandemic around the world. We use databases such as Bitcoinabuse\footnote{https://www.bitcoinabuse.com/} and Cryptoscamdb\footnote{https://cryptoscamdb.org/} where malicious tagged Bitcoin addresses are publicly available. As of $27^{th} Feb$ of $2021$, there were 54,036 maliciously tagged Bitcoin addresses and belong to types such as ``Scams'' and are reported by other users. However, in this work we do not consider the types of the malicious addresses rather consider them as malicious only.

\vspace{-0.3cm}
\subsection{Results Analysis}

For a clear understanding, we first answer RQ2 and then proceed to answer RQ1, RQ3, and RQ4. Note that we do not tweek any parameter whatsoever.

\subsubsection{RQ2: Are the features identified in state-of-the-art approaches targeting permissionless blockchains such as Ethereum applicable in Bitcoin or not?}\label{sub:RQ2}

Bitcoin has a different architecture than Ethereum. Due to the change-address concept, temporal features proposed in~\cite{shikar2020}, such as attractiveness and bursty behavior of in-degree and out-degree, are spread over multiple addresses. Applying ML techniques on the temporal features as identified in~\cite{shikar2020} in Bitcoin directly does not provide any reliable results. 

\begin{figure}
    \centering
    \includegraphics[width=\columnwidth]{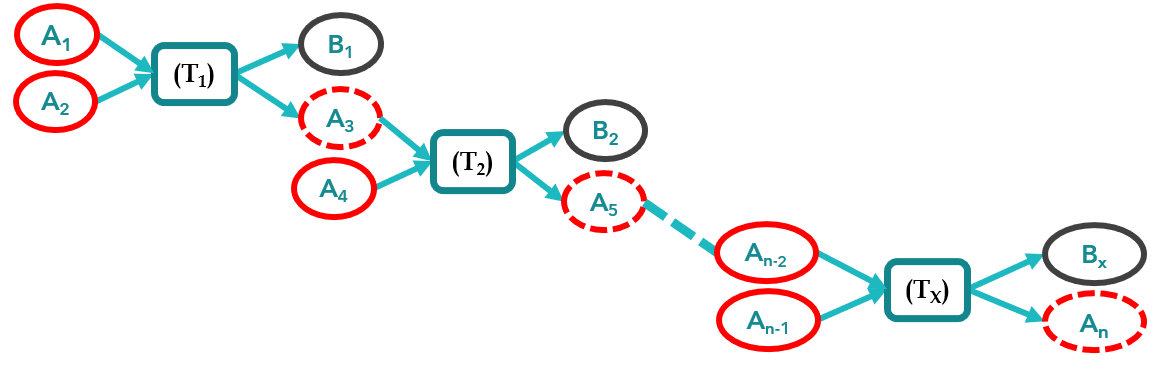}
    \caption{Example of user's out-degree being spread over his change addresses.}
    \label{fig:outdegree_spread_over_addresses}
    \vspace{-0.3cm}
\end{figure}

\begin{figure}
    \centering
        \includegraphics[width=0.7\columnwidth]{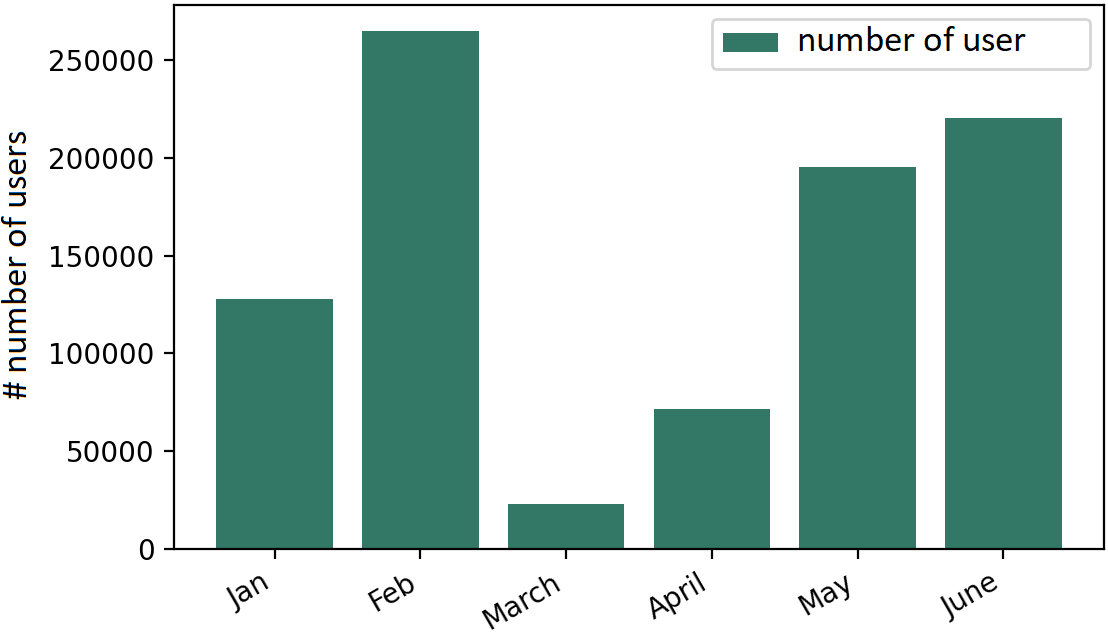}
        \caption{Number of users after address clustering.}
        \label{fig:clustered_addresses}
    \vspace{-0.3cm}
\end{figure}

\begin{figure*}
    \centering
     \vspace{-0.5cm}
    \subfloat[15Days.]
    [15Days.]{
        \includegraphics[width=0.32\textwidth]{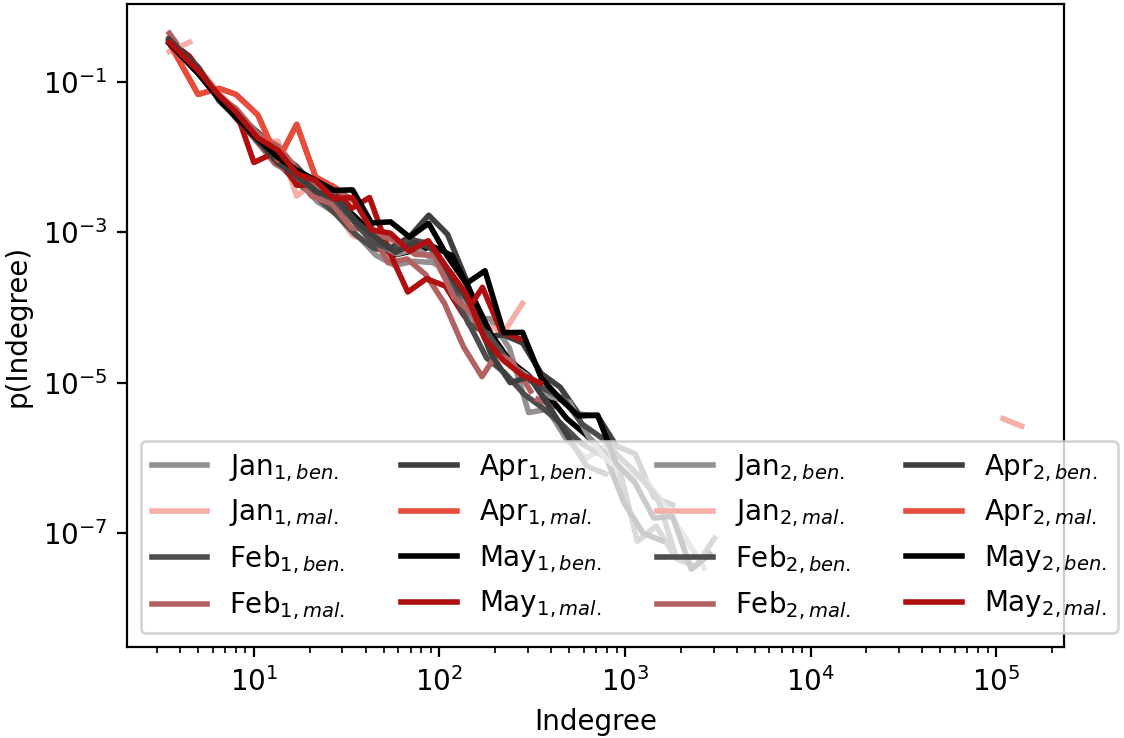}
        \label{fig:temporal_indegree_15days}
    }
    \subfloat[15Days.][15Days.]{
        \includegraphics[width=0.32\textwidth]{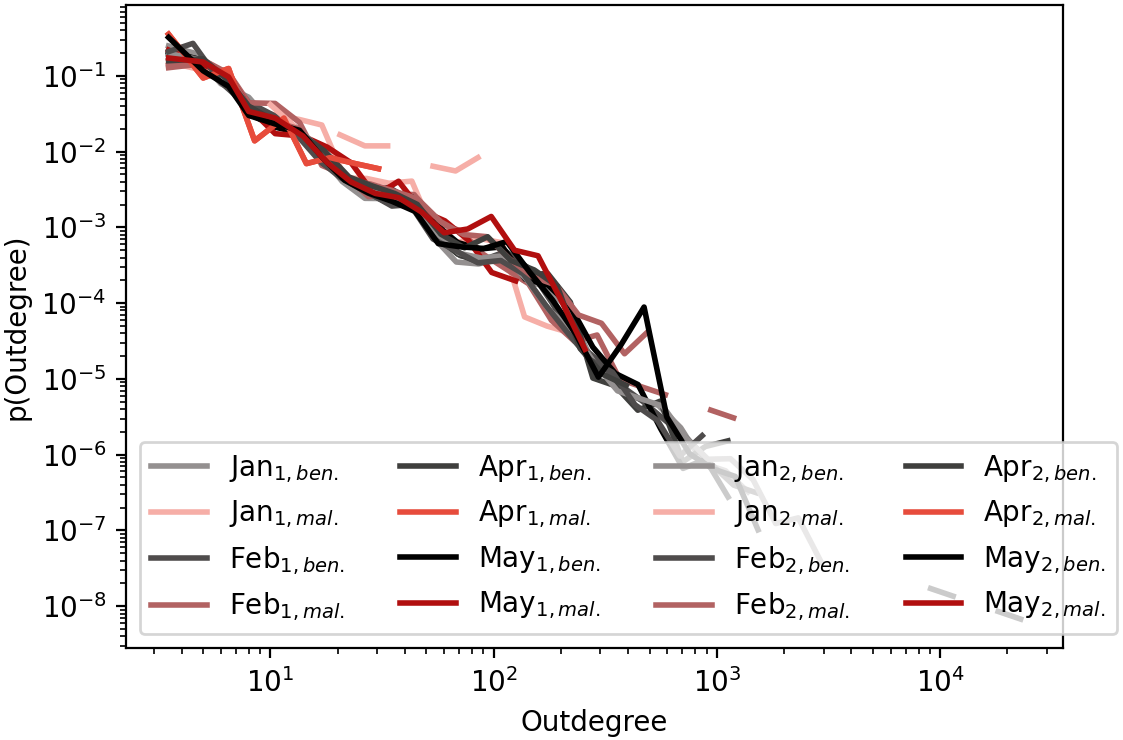}
        \label{fig:temporal_outdegree_15days}
    }
    \subfloat[15Days.][15Days.]{
        \includegraphics[width=0.32\textwidth]{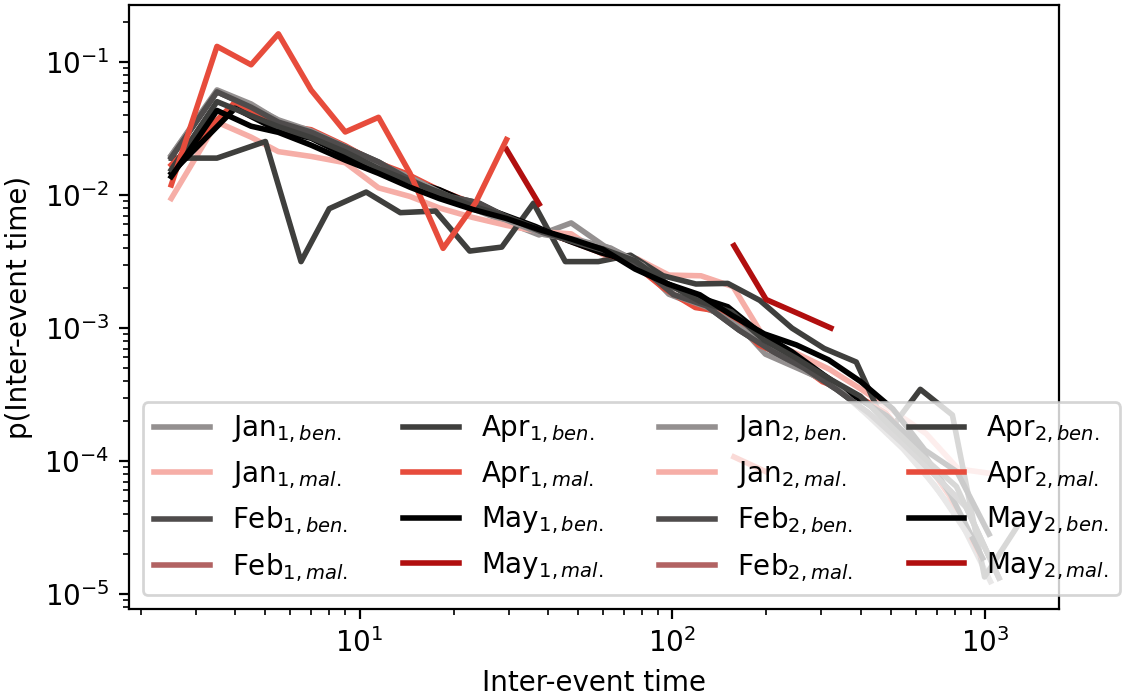}
        \label{fig:iet15}
    }\\
    \vspace{-0.2cm}
    \subfloat[1Month.][1Month.]{
        \includegraphics[width=0.32\textwidth]{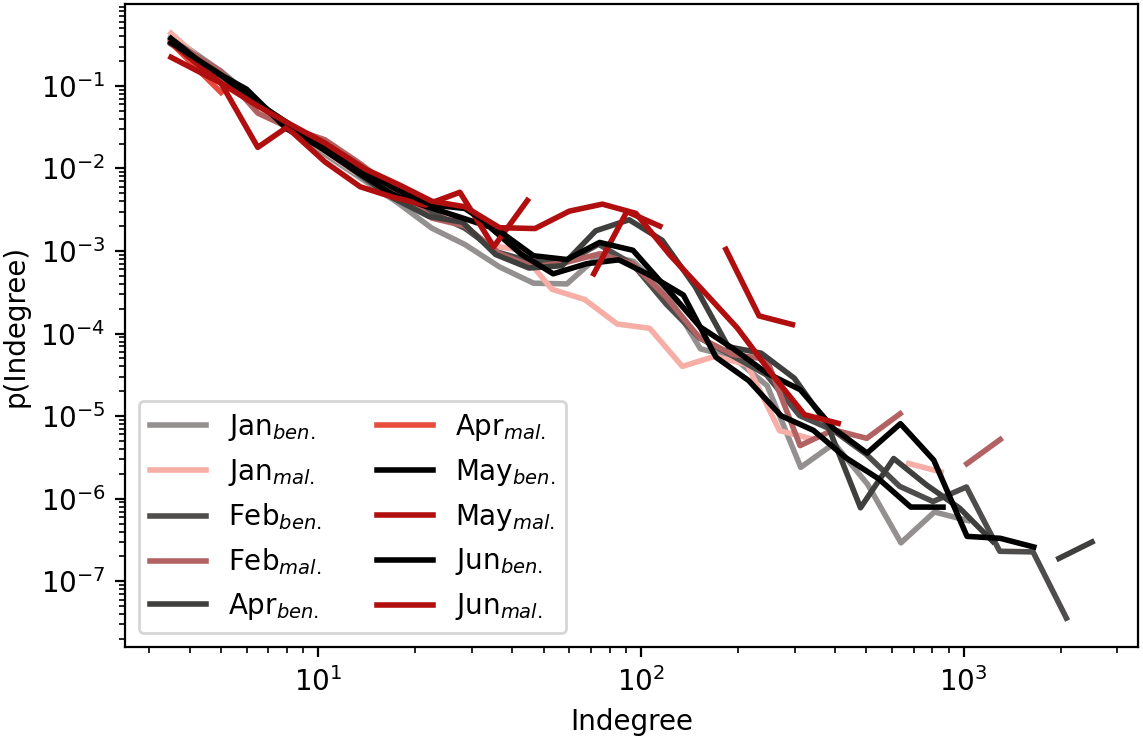}
        \label{fig:temporal_indegree_30days}
    }
    \subfloat[1Month.][1Month.]{
        \includegraphics[width=0.32\textwidth]{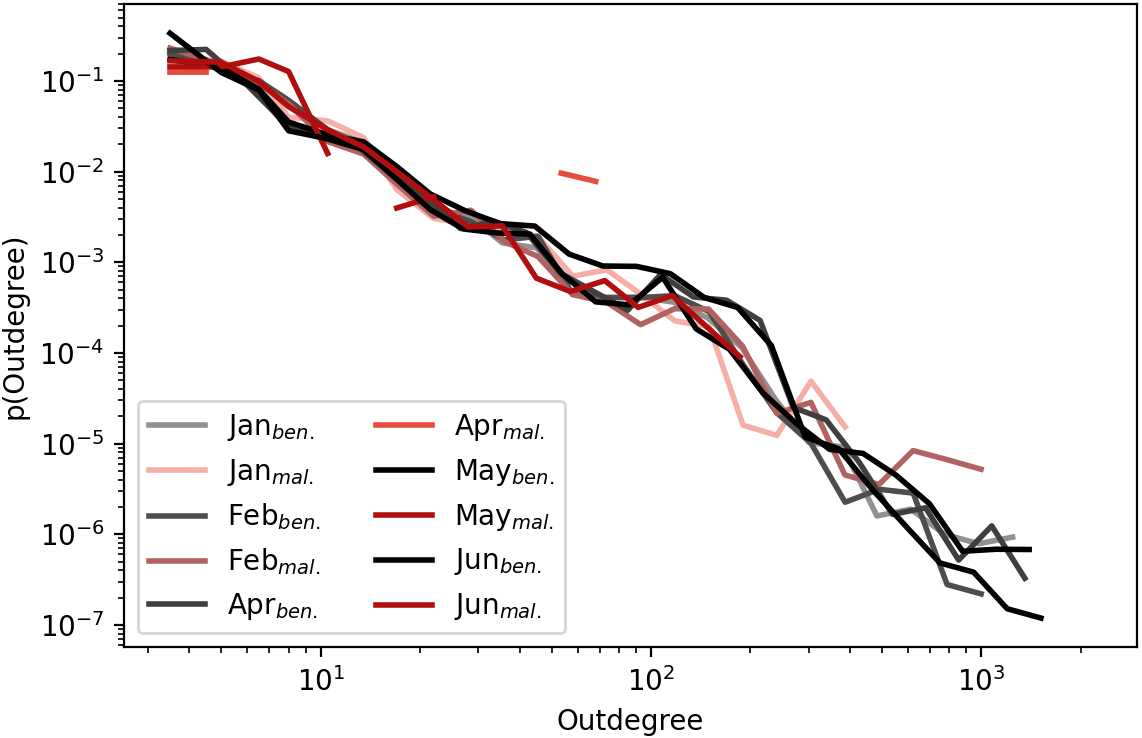}
        \label{fig:temporal_outdegree_30days}
    }
    \subfloat[1Month.][1Month.]{
        \includegraphics[width=0.32\textwidth]{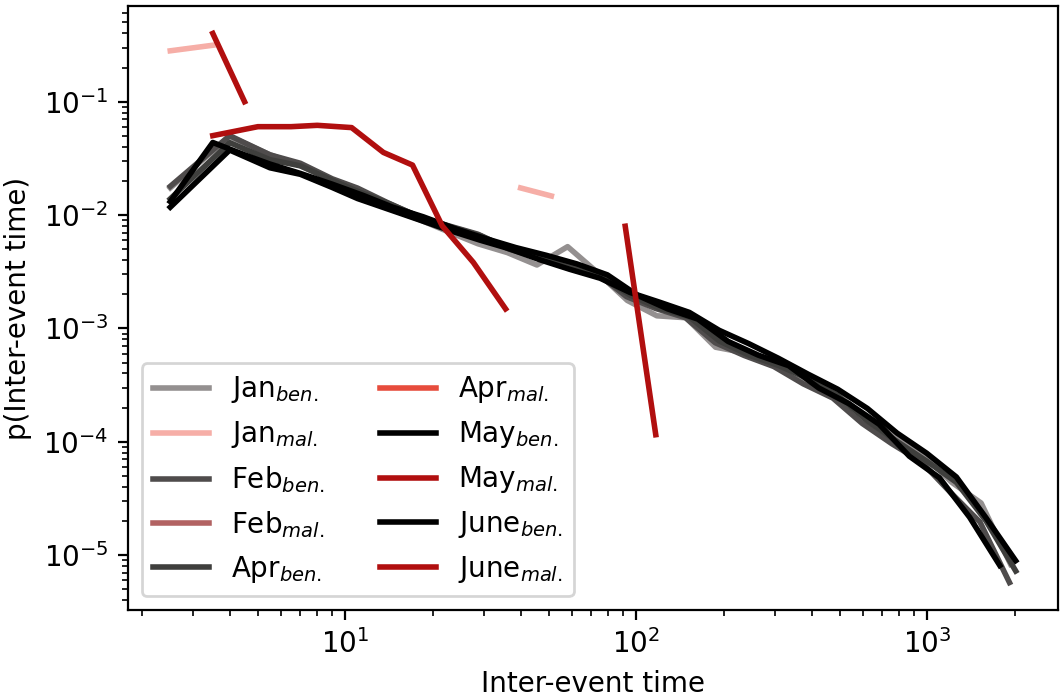}
        \label{fig:iet30}
    }
    \caption{Distributions, (a, d) in-degree, (b, e) out-degree, (c, f) inter-event time. Here, $_{1,ben.}$ represents benign addresses of first 15Days of a month, $_{2,ben.}$ represents benign addresses of second 15Days. Similarly $_{1,mal.}$ represents malicious addresses of first 15Days of a month, $_{2,mal.}$ represents malicious addresses of second 15Days. For (c, f) the x-axis is time between 2 transactions in units=blocks. While for (a, b, d, e) x-axis unit is number.} 
    \label{fig:distributions} 
    \vspace{-0.3cm}
\end{figure*}

From Figure~\ref{fig:outdegree_spread_over_addresses}, assume that addresses $A_1,A_2,A_3,\cdots,A_n$ (denoted as a red oval) belong to the same user $A$. Let these addresses perform $x$ number of transactions, $T_1,T_2,\cdots,T_x$, with addresses $\{B_1,B_2,\cdots, B_x\}$ (denoted as a black oval). Here, $\{A_3,A_5,\cdots,A_n\}$ are the change addresses of $A$ (denoted as a dotted oval). The out-degree of $A$ is spread over user's change addresses $A_3,A_5,\cdots,A_n$. Clustering such addresses give a temporal notion to the graph properties. If we do not cluster addresses, each address would be treated as different and temporal properties would not be correctly captured. Thus, the features identified in state-of-the-art approaches targeting permissionless blockchains such as Ethereum are not directly applicable in Bitcoin. In Bitcoin data, there exists few self-change addresses also (cf. Figure~\ref{fig:tempGranularityDistib}).

We first validate this using different SDs in $SD^{1}(T_g)$ as done in~\cite{shikar2020}. 
Here, we apply the K-Means algorithm with $K$=10 on the entire considered data. As a result, we get millions of suspicious addresses. However, tagging millions of the addresses as malicious is similar to tagging entire Bitcoin as malicious, which is impractical.

We now perform address clustering using multi-input and change address heuristics. Figure~\ref{fig:clustered_addresses} shows the number of unique users identified that performed a transaction every month. After address clustering, the total number of addresses that remain vary between 0.96\% and 1.75\% of the total number of addresses. Nonetheless, for June, the heuristics method cluster all the addresses as a single user. We do not know the exact cause for such a result. However, as one of the future investigations we would like to investigate the context and the reason behind such a result.

To understand whether the temporal features (such as burst and attractiveness) are applicable in Bitcoin or not, we generate distribution for in-degree and out-degree for both malicious and benign addresses. In~\cite{shikar2020}, the authors identified that \textit{power-law} distribution fits in-degree and out-degree in Ethereum. Here, from Figure~\ref{fig:distributions}, we find that power-law distribution also fits the in-degree and the out-degree for malicious and benign addresses present in Bitcoin. We find that, on average, power-law distribution with $x_{min}$=2.3, $\alpha \in$\{2.04, 2.06\} for malicious addresses and $\alpha$$\in$\{2.0, 1.9\} for benign addresses fits the in-degree in 15Days and 1Month temporal granularities, respectively (cf. Figure~\ref{fig:distributions}(a, d)). Here $\alpha$ and $x_{min}$ are the power-law exponents and the minimum $x$ from where the power-law distribution is observed. Further, on average, power-law distribution with $x_{min}$=2.3, $\alpha$$\in$\{1.88, 2.1.78\} for malicious addresses and $\alpha$$\in$\{1.84, 1.89\} for benign addresses fits the out-degree for different temporal granularities, respectively (cf. Figure~\ref{fig:distributions}(c, e)). To understand the difference between the two distributions identified using Bitcoin and Ethereum transaction data, we compute the \textit{KL-divergence} (KLD) between them. The KLD score between Ethereum and Bitcoin addresses for in-degree is \{0.051, 0.174\} for benign and malicious addresses, respectively. While the KLD for out-degree is \{0.124, 0.059\} for benign and malicious addresses. This means there is a minimal difference between the distributions of these features. Thus, showing that the addresses in two blockchains behaves similarly. 

In 15Days granularities, on average, (cf. Figure~\ref{fig:distributions}(c, f)), \textit{truncated power-law} with $x_{min}$=2.3, $\lambda$=\{0.00135, 0.00164\}, and $\alpha$=\{1.0002, 1.00008\} fits the inter-event time for benign and malicious addresses, respectively. Note that the truncation occurs at $\beta$=1/$\lambda$. In 1Month temporal granularity (cf. Figure~\ref{fig:iet30}), for benign addresses, the distribution of inter-event times fit truncated power-law with $\lambda$=0.0008, $\alpha$=1.00002, and $x_{min}$=2.3. However, for malicious addresses \textit{positive log-normal} distribution fits the best with $x_{min}$=2.3, $\mu$=5.088 and $\sigma$=1.737. The existence of power-law distribution indicates a bursty behavior both in in-degree, out-degree, and inter-event times of addresses. Hence, the features identified in state-of-the-art approaches targeting Ethereum are also applicable in Bitcoin but after change address clustering.

\begin{figure}
    \vspace{-0.3cm}
    \centering
    \subfloat[$SD^{2}(15Days)$]
    [$SD^{2}(15Days)$]{
        \includegraphics[width=0.48\columnwidth]{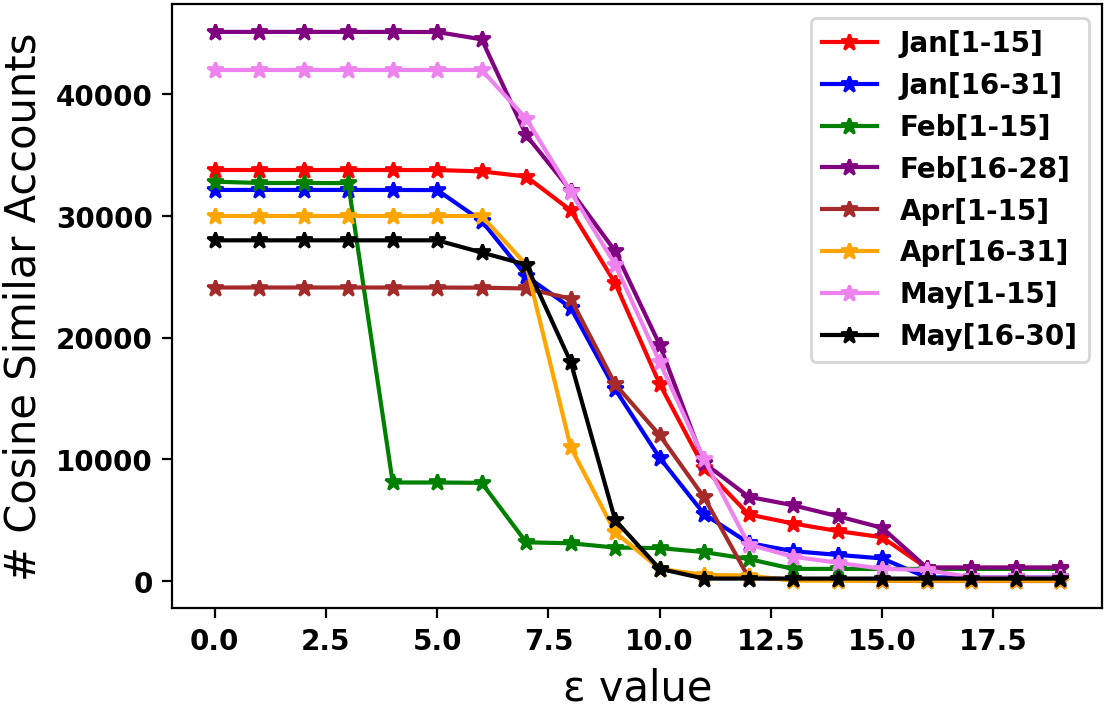}
        \label{cs_non_temp_15day}
    }
    \subfloat[$SD^{2}(1Month)$][$SD^{2}(1Month)$]{
        \includegraphics[width=0.48\columnwidth]{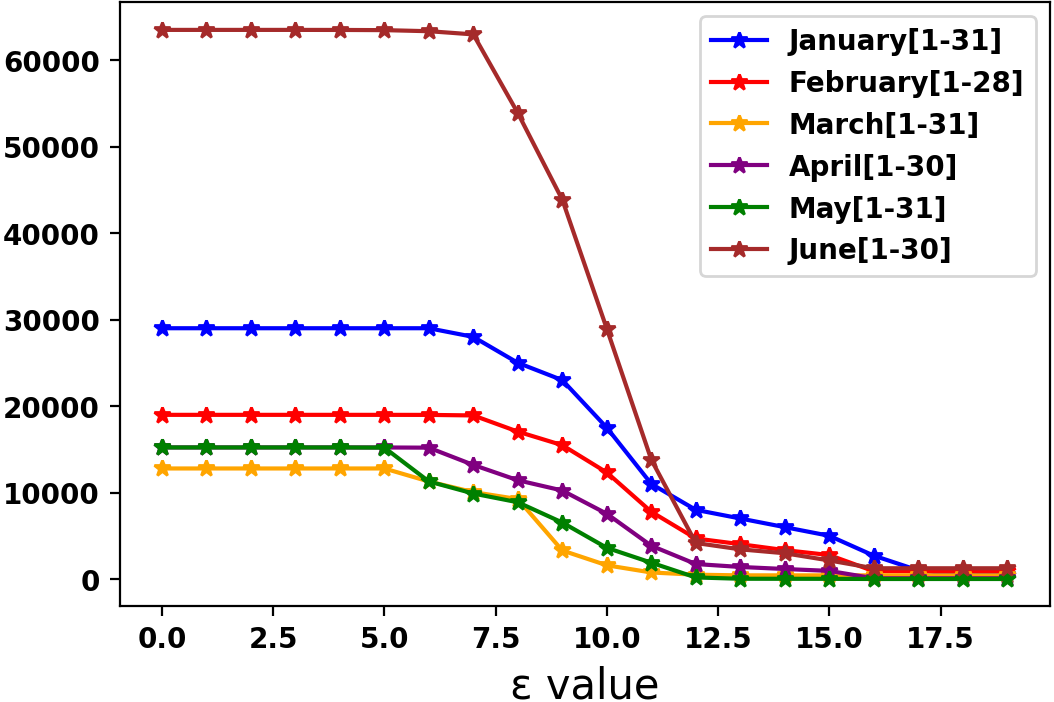}
        \label{cs_non_temp_30days}
    }\\
    \subfloat[$SD^{3}(15Days)$]
    [$SD^{3}(15Days)$]{
        \includegraphics[width=0.48\columnwidth]{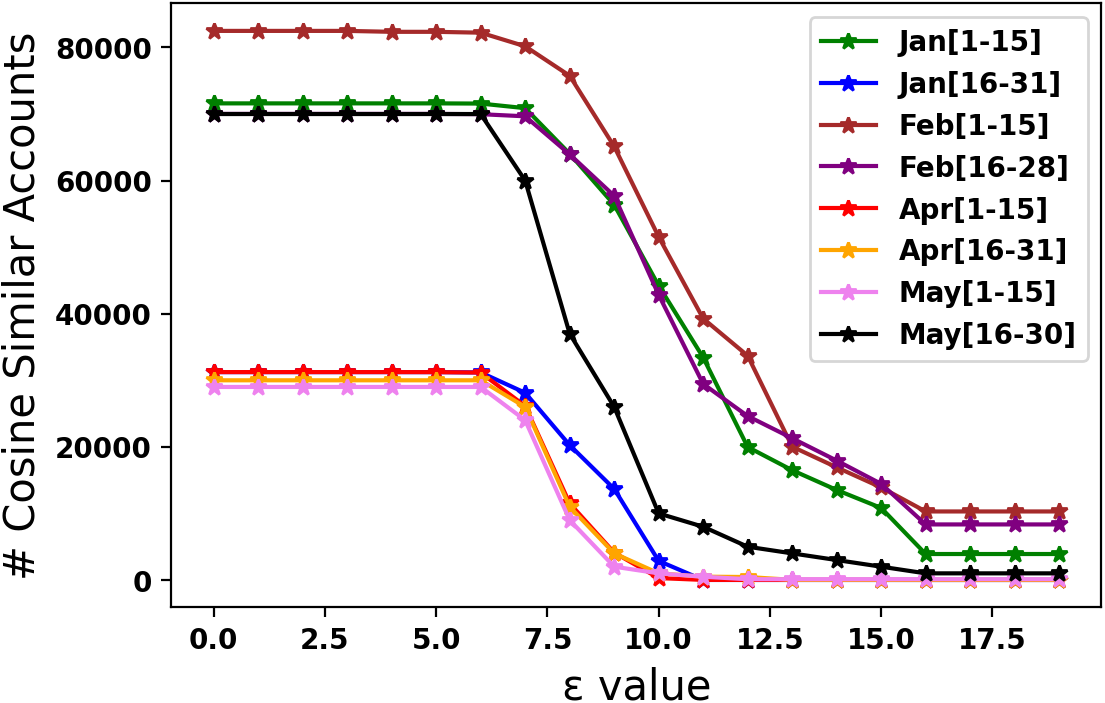}
        \label{cs_temp_15day}
    }
    \subfloat[$SD^{3}(1Month)$][$SD^{3}(1Month)$]{
        \includegraphics[width=0.48\columnwidth]{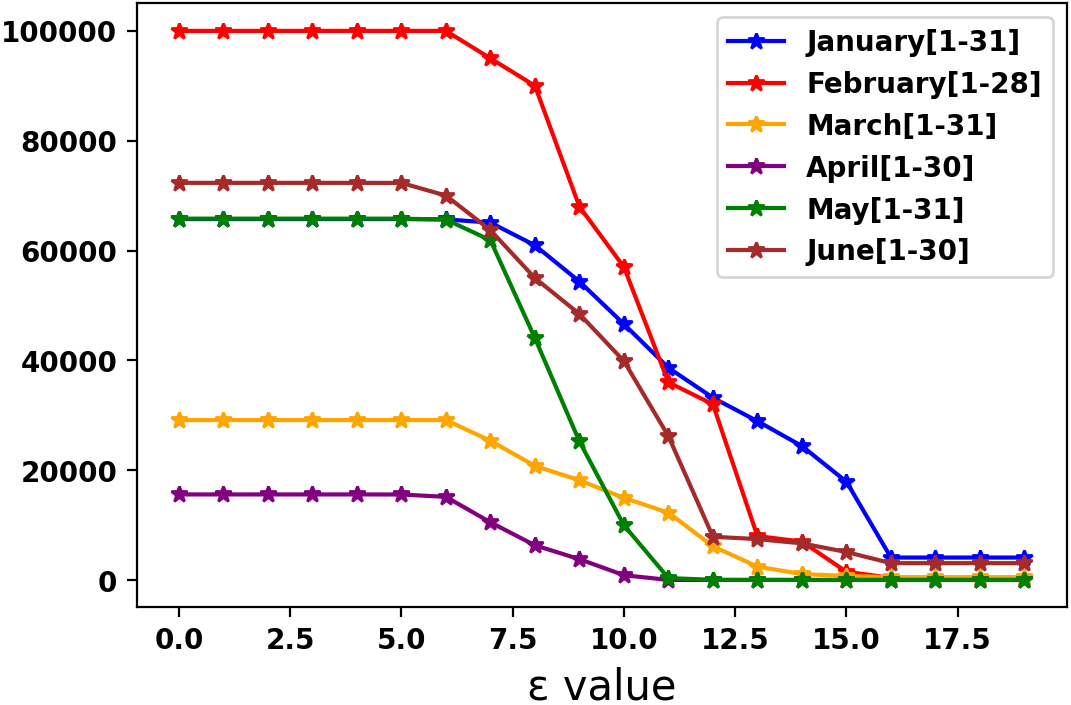}
        \label{cs_temp_30days}
    }
    \caption{Number of benign addresses having high CS with malicious addresses for different values of $\epsilon$.}\label{fig:cosine_similarity}
    \vspace{-0.3cm}
\end{figure}

\subsubsection{RQ1: Can we detect malicious addresses in Bitcoin using ML techniques and consider the temporal evolution of the graph?}

To answer this, we first apply K-Means with $K$=10 on all SDs in $SD^{2}(15Days)$ and $SD^{2}(1Month)$. On the clusters identified by the K-Means, we choose the cluster with maximum tagged malicious addresses and find the CS scores between malicious and benign addresses in that cluster. Figure~\ref{fig:cosine_similarity}(a, b) shows the number of benign addresses having high CS score obtained for different $\epsilon$. For SDs in $SD^{2}(15Days)$, we observe that $\epsilon$$\in$[7,16] provides acceptable results where all benign addresses are not marked as malicious but only a few highly similar addresses are marked. While for SDs in $SD^{2}(1Month)$, we observe that the $\epsilon$$\in$[12, 16]. Note that, in Figure ~\ref{fig:cosine_similarity}(a, c) as we did not had any significant number of malicious addresses, we could not compute the cluster with maximum number of malicious addresses for March for 15Days temporal granularity. While for June, all addresses clustered into one cluster.

Next, we extract all the temporal features introduced in~\cite{shikar2020} except the Ethereum blockchain specific features such as \textit{max gas price} and \textit{gas price burst}. After applying K-Means on all the SDs in $SD^{3}(15Days)$ and $SD^{3}(1Month)$, we perform similar steps as before. 

Figure~\ref{fig:cosine_similarity}(c, d) shows the number of high similarity addresses obtained for different $\epsilon$ for different sub-datasets when using both temporal and non-temporal features. For $SD^{3}(15Days)$, we observe that $\epsilon$$\in$[10, 16] provides acceptable results where all benign addresses are not marked as malicious but only a few highly similar addresses are marked. While for $SD^{3}(1Month)$, we observe that the $\epsilon$$\in$[11, 16]. 

\subsubsection{RQ3: What changes occur in the result after the change address clustering?}

From the analysis in RQ1, we find 87,907 and 68,215 malicious addresses in $SD^{3}(15Days)$ and $SD^{3}(1Month)$. While we detect 43,366 and 16,575 malicious addresses in $SD^{2}(15Days)$ and $SD^{2}(1Month)$. Here, we can see that there is significant increase in the number of suspect addresses after adding the temporal features. We also observe that 5,080 suspect addresses are common in $SD^{2}(15Days)$ and $SD^{3}(15Days)$. While 10,292 suspect addresses are common in $SD^{2}(1Month)$ and $SD^{3}(1Month)$.

We find that the detected malicious addresses in the period of 6 months performed 1,77,907 transactions. These transactions worth 25,824.326 BTC which is 0.01\% of total worth of benign transactions in that period. Thus, we verify that temporal features provide effective results towards detecting malicious addresses and identify changes that occur when we add temporal features. 

\subsubsection{RQ4: Does behavior change exists in Bitcoin addresses?}

Here, we study the distribution of the probability of an address $k$ to be malicious ($p^k$) using SDs in $SD^{2}(15Days)$, $SD^{2}(1Month)$,  $SD^{3}(15Days)$, and $SD^{3}(1Month)$. 

\begin{figure*}
    \centering
    \subfloat[15Days.]
    [15Days.]{
        \includegraphics[width=0.25\textwidth]{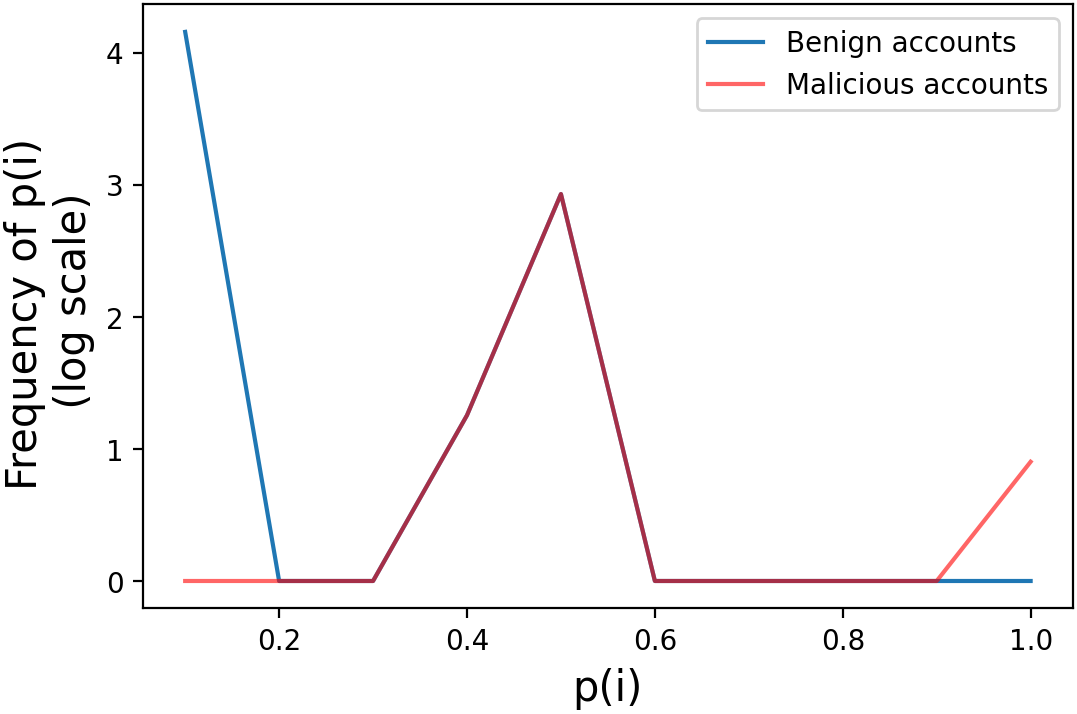}
        \label{fig:probability_non_temporal_15}
    }
    \subfloat[1Month.][1Month.]{
        \includegraphics[width=0.23\textwidth]{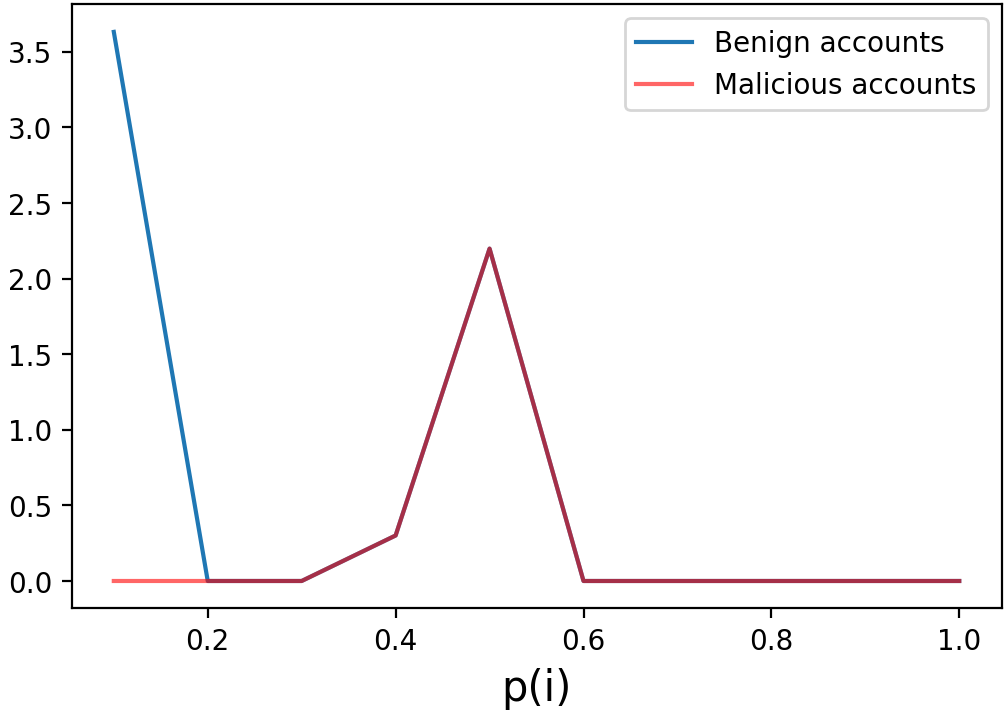}
        \label{fig:probability_non_temporal_30}
    }
    \subfloat[15Days.]
    [15Days.]{
        \includegraphics[width=0.23\textwidth]{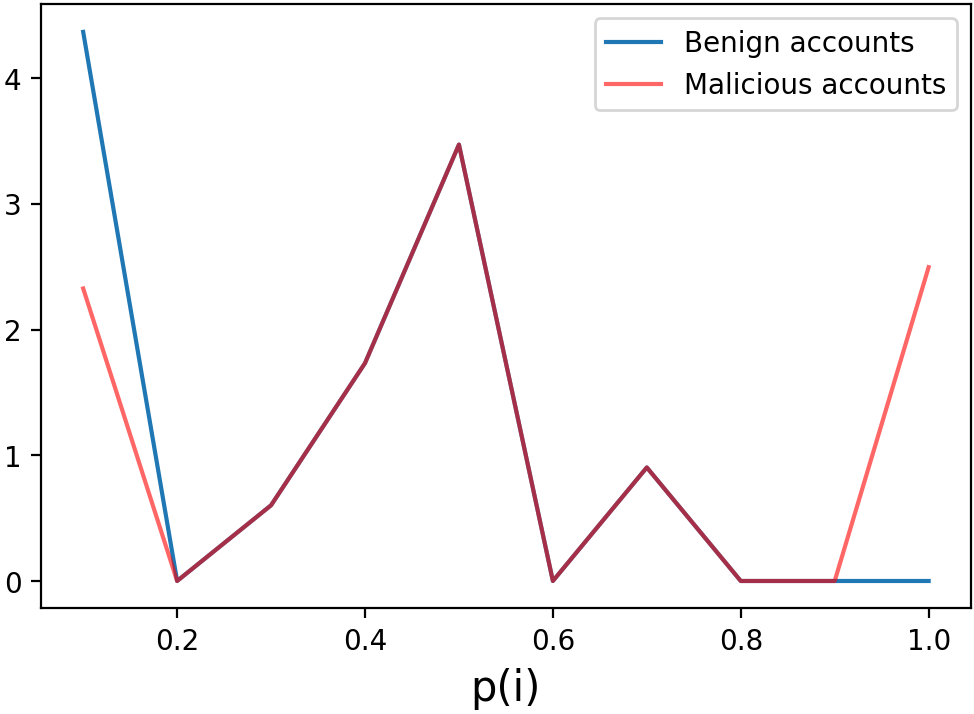}
        \label{fig:probability_temporal_15}
    }
    \subfloat[1Month.][1Month.]{
        \includegraphics[width=0.23\textwidth]{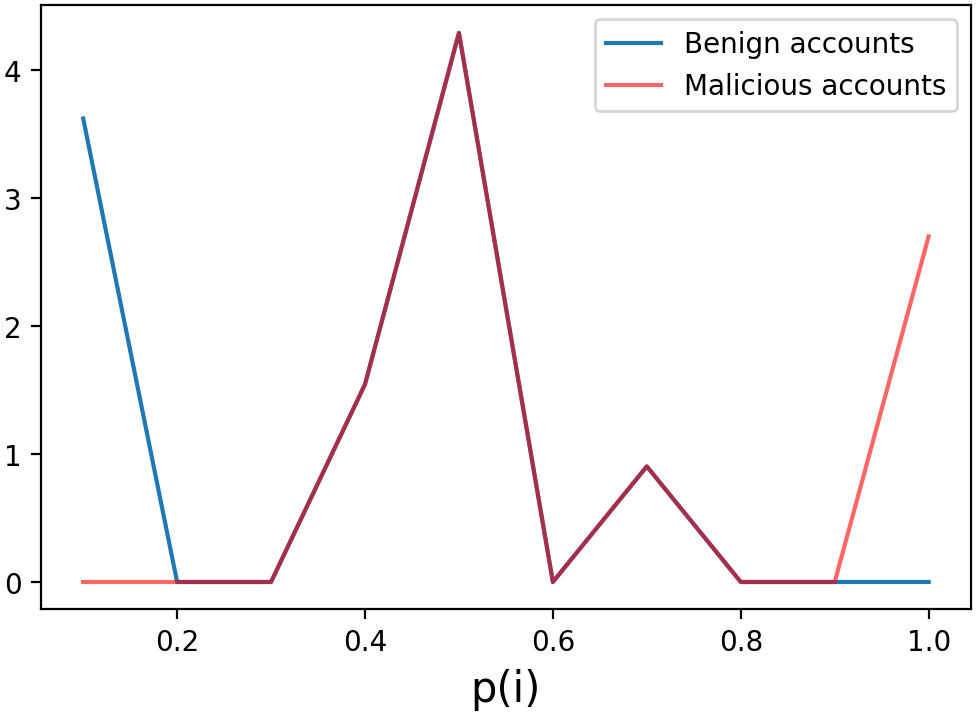}
        \label{fig:probability_temporal_30}
    }
    \caption{Distribution of $p^k$ for malicious and benign addresses.}\label{fig:probability} 
    \vspace{-0.3cm}
\end{figure*}

Figure~\ref{fig:probability}(a, b) shows the frequency of $p^k$ for benign addresses (blue line) and the frequency of $p^k$ for addresses detected as malicious (red line), using SDs in $SD^{2}(15Days)$ and $SD^{2}(1Month)$. As expected, for most benign addresses, we observe that their behavior does not change. While for few addresses there is a very high probability of being malicious. We also observe that some addresses change their behavior and their $p^k$$\in$[0.3,0.6]. We find that 868 addresses in $SD^{2}(15Days)$ change their behavior between malicious and benign. While 159 addresses in $SD^{2}(1Month)$ change their behavior between malicious and benign. Further, we find that 8 addresses using 15Days temporal granularity and 0 addresses using 1Month temporal granularity and have $p^k$=1.

Similar to the previous case, Figure~\ref{fig:probability}(c, d) shows the frequency of $p^k$ for benign address (blue line) and the frequency of $p^k$ for addresses detected as malicious (red line), using SDs in $SD^{3}(15Days)$ and $SD^{3}(1Month)$. In both the figures, we see a hike in the frequency of $p^k$ for both malicious and benign address when $p^k$$\geq$0.3. Further, as expected for the benign addresses, most of the addresses do not change their behavior, i.e., $p^k$=0. While, for some addresses $p^k$=1, which means that these addresses are highly suspicious. We find that 3,273 addresses in $SD^{3}(15Days)$ change their behavior between malicious and benign. While 19,712 addresses in $SD^{3}(1Month)$ change their behavior between malicious and benign. Further, we find that 313 addresses using 15Days temporal granularity and 501 addresses using 1Month temporal granularity and have $p^k$=1. We find that 3 addresses are detected by both 15Days and 1Month temporal granularity. We do not reveal the identity so as to not publicize these addresses as suspects.

Thus, we validate that behavior changes also exist in Bitcoin. More generally, the claim of~\cite{shikar2020} about the applicability of temporal features in any permissionless blockchain is valid in Bitcoin.
\vspace{-0.2cm}
\section{Conclusions and Future Work}\label{sec:Conclusion}

With an increase in the adoption of Bitcoin, security threats have become inevitable, more diverse, and are deployed using much advanced techniques. Currently, there is a limited work that focuses on detecting the addresses involved in malicious activities in Bitcoin, and those available do not consider the temporal aspects of transactions in Bitcoin. However, temporal aspects are well studied in other blockchain such as Ethereum. In this work, we consider the temporal nature of Bitcoin transactions to detect malicious addresses, study their behavior, and validate the state-of-the-art methods for Bitcoin. 

We observe that change-address heuristics plays an important role in extraction of temporal features, the state-of-the-art temporal features have similar behavior in Bitcoin as well, and temporal features are effective in terms of detecting more suspects. We find, with high probability, 3 suspect addresses that were detected across different temporal granularities. 
In the future, we would like to test the state-of-the-art method with more data to provide more robust results. Bitcoin is an extremely large connected graph with more than 651 Million transactions over 11 years. Not utilizing complete data might induce bias in the study and might not capture the true behavior. In our case, the application of heuristics, the results might have a bias probably which we see in month of June. If the heuristics algorithm provides the best results, our results would be best. However, we can not guarantee the effectiveness of the heuristics approach as we lack the ground truth about the addresses held by an entity. We would like to study the effectiveness of the heuristics algorithm in future.
\vspace{-0.3cm}
\section*{Acknowledgements}
This work is partially funded by the National Blockchain Project (grant number NCSC/CS/2017518) at IIT Kanpur sponsored by the National Cyber Security Coordinator's office of the Government of India and partially by the C3i Center funding from the Science and Engineering Research Board of the Government of India (grant number SERB/CS/2016466).
\vspace{-0.3cm}
\bibliographystyle{plain}
\bibliography{citations}
\end{document}